\documentclass[paper=a4, fontsize=11pt]{scrartcl}
\usepackage{graphicx}
\usepackage{amssymb}
\begin{document}
\begin{center}{\large{Einstein, Weyl and Asymmetric Geometries}}
\end{center}
\vspace*{1.5cm}
\begin{center}
A. C. V. V. de Siqueira
$^{*}$ \\
Retired \\
Universidade Federal Rural de Pernambuco \\
52.171-900, Recife, PE, Brazil.\\
${}^*$ E-mail:antonio.vsiqueira@ufrpe.br
\end{center}
\vspace*{1.5cm}
\begin{center}
{ Abstract}
\end{center}
\vspace*{1.0cm}
\begin{center}
In a previous paper, we presented new results on non-Riemannian geometry. For an asymmetric connection, we showed that a projective change in the symmetric part generates a vector field that is not arbitrary, but is the gradient of a non-arbitrary scalar field. As a consequence, Weyl's geometry is a conformal differential geometry and is associated with asymmetric geometry by this projective change. In the present paper, important differences between light-like and non-light-like intervals are analysed. We show that the integrability condition in Weyl’s geometry, together with the condition that null and massive particles interact locally in an Einstein spacetime, implies microscopic spacetime oscillations of Weyl’s and Einstein’s geometries. We construct an equation for linear transverse waves and make a qualitative application in a hydrogen atom, which, from qualitative viewpoint, agrees with Bohr's hydrogen model and Schrodinger's quantum mechanics.
\textbf{•}
\end{center}
\newpage
\section{Introduction}
In this paper, important differences between light-like and non-light-like intervals are analysed. 
\newline
It is also important to note that Weyl's geometry and asymmetric geometry depend on the scalar field $\psi$ \cite{1}, but Weyl's projective tensor and Weyl's conformal tensor do not, because they were constructed for this purpose. There are other tensors in asymmetric geometry that also do not depend on $\psi$ \cite{2}. Although important, none of these tensors will be considered in this paper.
\newline
As a consequence of the integrability condition in Weyl’s geometry, together with the condition that null and massive particles interact locally in an Einstein spacetime, oscillatory scalar fields generate microscopic oscillations in spacetime in Weyl’s and Einstein’s geometries. A classical transverse wave equation is obtained for a flat spacetime.
\newline
This paper is organized as follows: In Sec. $2$, we present some elements of two geometries and two equations that govern scalar fields, one associated to asymmetric geometry and the other associated to Weyl's geometry, \cite{1}. In Sec. $3$, we use the method of change of variables to solve two Linear First-Order Partial Differential Equations that govern the scalar fields associated with asymmetric geometry and Weyl's geometry, \cite{3}. In Sec. $4$, we analyse the integrability condition in Weyl's geometry. In Sec. $5$, we use the oscillatory solutions of these scalar fields, interpret these solutions as microscopic oscillations of a flat spacetime and build a classical wave equation for a particle. In Sec. $6$, we present our concluding remarks. 
\newpage
\renewcommand{\theequation}{\thesection.\arabic{equation}}
\section{Asymmetric Connection and Weyl's Connection}
\setcounter{equation}{0}
 $         $
This section is based on a previous paper \cite{1}. To emphasize differences between like-light and not like-light intervals, some notations used for scalar and vector fields will be changed.
\newline
We define an asymmetric connection as follows
\begin{equation}
L_{\mu \nu}^{\eta}=\Gamma_{\mu \nu}^{\eta}+\Omega_{\mu \nu}^{\eta},
\end{equation} 
\begin{equation}
 \Gamma_{\mu \nu}^{\eta} =\Gamma_{\nu \mu}^{\eta},
\end{equation}
and
\begin{equation}
\Omega_{\nu \mu}^{\eta}=-\Omega_{\mu \nu}^{\eta}.
\end{equation}
Let us consider two vector fields with the same direction at each point where its components satisfy the following relation
\begin{equation}
\widehat{X}^{\eta}=\varphi X^{\eta}.
\end{equation}
As a consequence of (2.4), for a projective change of connection, we get 
\begin{equation}
{\widehat{L}}_{\mu \nu}^{\sigma}={L}_{\mu \nu}^{\sigma}+2{\delta}_{\mu}^{\sigma}{\psi}_{\nu},
\end{equation}
\begin{equation}
\widehat{\Gamma}_{\mu \nu}^{\eta}=\Gamma_{\mu \nu}^{\eta}+{\delta}_{\mu}^{\eta}{\psi}_{\nu}+{\delta}_{\nu}^{\eta}{\psi}_{\mu},
\end{equation} 
and
\begin{equation}
\widehat{\Omega}_{\mu \nu}^{\eta}=\Omega_{\mu \nu}^{\eta}+{\delta}_{\mu}^{\eta}{\psi}_{\nu}-{\delta}_{\nu}^{\eta}{\psi}_{\mu}.
\end{equation} 
Using normal coordinates, $y^{\eta}$ and $z^{\eta}$, for a projective change of connection, we have
\begin{equation}
 y^{\eta}=\frac{z^{\eta}}{f(z)}. 
\end{equation}
As a consequence, 
\begin{equation}
2\psi_{\mu}=\partial_{\mu}log(f^{-1}+\frac{\partial{f^{-1}}}{\partial{z^{\sigma}}}z^{\sigma}).
\end{equation}
We now define
\begin{equation}
f^{-1}=h,
\end{equation}
then
\begin{equation}
h+\frac{\partial{h}}{\partial{z^{\sigma}}}z^{\sigma}=e^{2\psi}.
\end{equation}
Weyl chose the following symmetric connection
\begin{equation}
{W}_{\mu \nu}^{\eta}=C_{\mu \nu}^{\eta}+A_{\mu \nu}^{\eta},
\end{equation}
and imposed the following condition to the metric tensor 
\begin{equation}
 g_{\mu \nu;\sigma}+2g_{\mu \nu}{\omega}_{\sigma}=0.
\end{equation}
In (2.12), $C_{\mu \nu}^{\eta}$ are components of the Christoffel symbols and $A_{\mu \nu}^{\eta}$ is a tensor with symmetric covariant indices. 
\newline
Using (2.13) in (2.12), we can show that
\begin{equation}
A_{\mu \nu}^{\eta}={\delta}_{\mu}^{\eta}{\omega}_{\nu}+{\delta}_{\nu}^{\eta}{\omega}_{\mu}-g_{\mu \nu}{\omega}^{\eta}.
\end{equation}
Replacing (2.14) in (2.12), we have
\begin{equation}
{W}_{\mu \nu}^{\eta}=C_{\mu \nu}^{\eta}+{\delta}_{\mu}^{\eta}{\omega}_{\nu}+{\delta}_{\nu}^{\eta}{\omega}_{\mu}-g_{\mu \nu}{\omega}^{\eta}.
\end{equation}
Let us call ${\omega}_{\nu}$ as ${\omega}^{(1)}_{\nu}$ for a non-light-like interval and ${\omega}^{(2)}_{\nu}$
for a light-like interval. 
\newline
We will then have
\begin{equation}
{W}_{\mu \nu}^{\eta}=C_{\mu \nu}^{\eta}+{\delta}_{\mu}^{\eta}\omega^{(1)}_{\nu}+{\delta}_{\nu}^{\eta}\omega^{(1)}_{\mu}-g_{\mu \nu}g^{\eta\sigma}\omega^{(1)}_{\sigma},
\end{equation}
for a non-light-like interval and
\begin{equation}
{W}_{\mu \nu}^{\eta}=C_{\mu \nu}^{\eta}+{\delta}_{\mu}^{\eta}\omega^{(2)}_{\nu}+{\delta}_{\nu}^{\eta}\omega^{(2)}_{\mu}-g_{\mu \nu}g^{\eta\sigma}\omega^{(2)}_{\sigma},
\end{equation}
for a light-like interval.
\newline
From $ g_{\mu \nu}{z}^{\mu}{z}^{\nu}\neq0 $ and a similar development used in obtaining (2.9), we get 
\begin{equation}
{\omega}^{(1)}_{\mu}=-2\partial_{\mu}logf_{(1)}+\partial_{\mu}log(f_{(1)}-\frac{\partial{f_{(1)}}}{\partial{z^{\sigma}}}z^{\sigma}).
\end{equation}
Using (2.10)) in (2.16), we have
\begin{equation}
h_{(1)}+\frac{\partial{h_{(1)}}}{\partial{z^{\sigma}}}z^{\sigma}=e^{{\omega}^{(1)}}.
\end{equation}
Let us consider a light-like interval, $z_{\sigma}z^{\sigma}=0$. 
\newline
After some development, we get
\begin{equation}
2{\omega^{(2)}}_{\mu}=\partial_{\mu}log(f_{(2)}^{-1}+\frac{\partial{f_{(2)}^{-1}}}{\partial{z^{\sigma}}}z^{\sigma}).
\end{equation}
As before, using
\begin{equation}
f^{-1}=h,
\end{equation}
we have
\begin{equation}
h_{(2)}+\frac{\partial{h_{(2)}}}{\partial{z^{\sigma}}}z^{\sigma}=e^{2{\omega^{(2)}}}.
\end{equation}
Then, for a non-light-like interval, a conformal transformation in Weyl's geometry induces a projective change in asymmetric geometry if 
\begin{equation}
2\psi=\omega^{(1)},
\end{equation}
in which
\begin{equation}
{W}_{\mu \nu}^{\eta}=C_{\mu \nu}^{\eta}+{\delta}_{\mu}^{\eta}\partial_{\nu}\omega^{(1)}+{\delta}_{\nu}^{\eta}\partial_{\mu}\omega^{(1)}-g_{\mu \nu}g^{\eta\sigma}\partial_{\sigma}\omega^{(1)},
\end{equation}
\begin{equation}
h_{(1)}+\frac{\partial{h_{(1)}}}{\partial{z^{\sigma}}}z^{\sigma}=e^{{\omega}^{(1)}},
\end{equation}
\begin{equation}
2\widehat{\Gamma}_{\mu \nu}^{\eta}=2\Gamma_{\mu \nu}^{\eta}+ +{\delta}_{\mu}^{\eta}\partial_{\nu}\omega^{(1)}+{\delta}_{\nu}^{\eta}\partial_{\mu}\omega^{(1)},
\end{equation}
\begin{equation}
2\widehat{\Omega}_{\mu \nu}^{\eta}=2\Omega_{\mu \nu}^{\eta}+{\delta}_{\mu}^{\eta}\partial_{\nu}\omega^{(1)}-{\delta}_{\nu}^{\eta}\partial_{\mu}\omega^{(1)}.
\end{equation} 
For a light-like interval, a conformal transformation in Weyl's geometry induces a projective change in asymmetric geometry if 
\begin{equation}
\psi=\omega^{(2)},
\end{equation}
in which
\begin{equation}
{W}_{\mu \nu}^{\eta}=C_{\mu \nu}^{\eta}+{\delta}_{\mu}^{\eta}\partial_{\nu}\omega^{(2)}+{\delta}_{\nu}^{\eta}\partial_{\mu}\omega^{(2)}-g_{\mu \nu}g^{\eta\sigma}\partial_{\sigma}\omega^{(2)},
\end{equation}
\begin{equation}
h_{(2)}+\frac{\partial{h_{(2)}}}{\partial{z^{\sigma}}}z^{\sigma}=e^{2\omega^{(2)}},
\end{equation}
\begin{equation}
\widehat{\Gamma}_{\mu \nu}^{\eta}=\Gamma_{\mu \nu}^{\eta}+{\delta}_{\mu}^{\eta}\partial_{\nu}\omega^{(2)}+{\delta}_{\nu}^{\eta}\partial_{\mu}\omega^{(2)},
\end{equation}
\begin{equation}
\widehat{\Omega}_{\mu \nu}^{\eta}=\Omega_{\mu \nu}^{\eta}+{\delta}_{\mu}^{\eta}\partial_{\nu}\omega^{(2)}-{\delta}_{\nu}^{\eta}\partial_{\mu}\omega^{(2)}.
\end{equation}
The connections have different geodesics and curvatures for different types of intervals.
\newline
This produces different effects in both Weyl's geometry and asymmetric geometry. In other words, we have one Weyl's geometry for light-like intervals and another Weyl's geometry for non-like-light intervals. The same is true for the asymmetric connection, as can be seen from (2.26), (2.27), (2.31) and (2.32).
\newline
In section three, we will see an infinite number of solutions for each function $\omega^{(1)}$ in equation (2.19) and $\omega^{(2)}$ (2.22). 
\newline
For an infinite subset of solutions of equations (2.24), (2.25), (2.24), (2.25), (2.26), (2.27), (2.29), (2.30), (2.31) and (2.32) with the following condition
\begin{equation}
2\psi=\omega^{(1)}=2\omega^{(2)},
\end{equation}
we have a unique Weyl's geometry and a unique asymmetric geometry, where null particles and massive particles can interact. In this case, Weyl's geometry is the same as Einstein's geometry.
\newline
For assymmetric geometry, using (3.1), (3.2), (3.3), (3.5), (3.6), (3.17) from reference \cite{1}, we have
\begin{equation}
\varphi=e^{-2\psi}.
\end{equation}
In Weyl's geometry, for a non-light-like interval, we get
\begin{equation}
\varphi=e^{-\omega^{(1)}},
\end{equation} 
and, for a light-like interval, 
\begin{equation}
\varphi=e^{-2\omega^{(2)}}.
\end{equation} 
Then, in Weyl's geometry, for a non-light-like interval, (2.4) will be
\begin{equation}
\widehat{X}^{\eta}=e^{-\omega^{(1)}}X^{\eta},
\end{equation}
and, for a light-like interval, 
\begin{equation}
\widehat{X}^{\eta}=e^{-2\omega^{(2)}}X^{\eta}.
\end{equation}
In Weyl's geometry for a non-light-like interval, we have 
 \begin{equation}
 G^{(1)}_{\mu\nu}=e^{\omega^{(1)}}g_{\mu\nu},
\end{equation} 
which is equivalent to
\begin{equation}
 g_{\mu \nu;\sigma}+g_{\mu \nu}\partial_{\mu}\omega^{(1)}=0,
\end{equation} 
and, for a light-like interval, 
\begin{equation}
 G^{(2)}_{\mu\nu}=e^{2\omega^{(2)}}g_{\mu\nu},
\end{equation} 
and
\begin{equation}
 g_{\mu \nu;\sigma}+2g_{\mu \nu}\partial_{\sigma}\omega^{(2)}=0.
\end{equation} 
Eliminating $g_{\mu \nu}$ in (2.39) and (2.41), we have
\begin{equation}
 G^{(1)}_{\mu\nu}=e^{\omega^{(1)}}e^{-2\omega^{(2)}}G^{(2)}_{\mu\nu}.
\end{equation} 
Let us consider
\begin{equation}
 \omega^{(1)}-2\omega^{(2)}\neq0,
\end{equation} 
then we can repeat the same procedures used for (2.39) and (2.41) an infinite number of times for each pair of functions $\omega^{(1)}$ and $\omega^{(2)}$ and write   
\begin{equation}
 G^{(2)}_{\mu\nu;\sigma}+2G^{(2)}_{\mu\nu}\partial_{\sigma}(\omega^{(1)}-2\omega^{(2)})=0.
\end{equation}
Note that light-like and non-light-like intervals are not the same in this new Weyl's geometry and asymmetric geometry, like those associated to (2.39) and (2.41).
\newline
We now consider an association among the Einstein, Weyl and asymmetric geometries.
\newline
Let us write Einstein's equations and Weyl's curvature tensor
\begin{equation}
 R_{\mu\nu}-\frac{1}{2}g_{\mu\nu}R=-const.T_{\mu\nu},
\end{equation}
\begin{equation}
\Sigma^{\alpha}{}_{\mu \sigma \nu }=-\partial_{\nu} W_{\mu \sigma
}^{\alpha}+\partial_{\sigma} W_{\mu \nu}^{\alpha}-W_{\mu
\sigma}^{\eta} W_{n \nu}^{\alpha}+W_{\mu \nu}^{\eta} W_{\sigma \eta}^{\alpha}.
\end{equation}
For Weyl's curvature tensor, we use the following contraction
\begin{equation}
 \Sigma^{\lambda}_{\mu\nu\lambda}= \Sigma_{\mu\nu}.
\end{equation}
Let us define  $\Delta_{1}$,  $\Delta_{2}$ and $\omega_{\mu\nu}$ as follows
\begin{equation}
 \Delta_{1}\omega= g^{\mu\nu}\partial_{\mu}\omega\partial_{\nu}\omega,
\end{equation}
\begin{equation}
 \Delta_{2}\omega= g^{\mu\nu}(\partial^{2}_{\mu\nu}\omega-C_{\mu \nu}^{\sigma}\partial_{\sigma}\omega),
\end{equation}
\begin{equation}
\omega_{\mu\nu}=(\partial^{2}_{\mu\nu}\omega-C_{\mu \nu}^{\sigma}\partial_{\sigma}\omega)-\partial_{\mu}\omega\partial_{\nu}\omega.
\end{equation}
Therefore, Einstein's tensor and Weyl's tensor are related by \cite{4},
\begin{equation}
 \Sigma_{\mu\nu}-\frac{1}{2}G_{\mu\nu}\Sigma-2\omega_{\mu\nu}+g_{\mu\nu}(\Delta_{1}\omega+2\Delta_{2}\omega)=R_{\mu\nu}-\frac{1}{2}g_{\mu\nu}R=-const.T_{\mu\nu}.
\end{equation}
Using (2.39) in (2.52), we have
\begin{eqnarray}
\nonumber \Sigma^{(1)}_{\mu\nu}-\frac{1}{2}e^{\omega^{(1)}}g_{\mu\nu}\Sigma^{(1)}-2\omega^{(1)}_{\mu\nu}+g_{\mu\nu}(\Delta_{1}\omega^{(1)}+2\Delta_{2}\omega^{(1)})\\
\nonumber =R_{\mu\nu}-\frac{1}{2}g_{\mu\nu}R=-const.T_{\mu\nu}.\\
\end{eqnarray}
Placing (2.41) in (2.52), we get
\begin{eqnarray}
\nonumber \Sigma^{(2)}_{\mu\nu}-\frac{1}{2}e^{2\omega^{(2)}}g_{\mu\nu}\Sigma^{(2)}-2\omega^{(2)}_{\mu\nu}+g_{\mu\nu}(\Delta_{1}\omega^{(2)}+2\Delta_{2}\omega^{(2)})\\
\nonumber =R_{\mu\nu}-\frac{1}{2}g_{\mu\nu}R=-const.T_{\mu\nu}.\\
\end{eqnarray}
From the mathematical viewpoint, equations (2.53) and (2.54) can map the same or two different space-time regions of general relativity in two different Weyl's geometries. However, as we shall see below, Einstein's geometry and integrability conditions in Weyl's geometry impose the condition that massive particles live in (2.53) (non-light-like interval) and null particles live in (2.54) (light-like interval). 
\newline
If we consider only massive or only null particles in Weyl's geometry, we have
\begin{equation}
 \omega^{(1)}-2\omega^{(2)}\neq0.
\end{equation} 
If we consider massive and null particles interacting in Weyl's geometry, then we need
\begin{equation}
 \omega^{(1)}-2\omega^{(2)}=0,
\end{equation} 
and Weyl's geometry will be an Einstein's geometry. 
\newline
Let us now consider asymmetric connections. 
\newline
For (2.2) and (2.3), we have the respective curvatures
\begin{eqnarray}
  \nonumber \widehat{B}^{\alpha}_{\mu \sigma \nu }=B^{\alpha}_{\mu \sigma \nu}+\delta^{\alpha}_{\nu}[\partial^{2}_{\mu \sigma}\psi-\Gamma_{\mu \sigma}^{\eta}\partial_{\eta}\psi-{\partial_{\mu}\psi}{\partial_{\sigma}\psi}]\\
  \nonumber -\delta^{\alpha}_{\sigma}[\partial^{2}_{\mu \nu}\psi-\Gamma_{\mu \nu}^{\eta}\partial_{\eta}\psi-{\partial_{\mu}\psi}{\partial_{\nu}\psi}],\\
\end{eqnarray}
and
\begin{eqnarray}
  \nonumber \widehat{ \Omega}^{\alpha}_{\mu \sigma \nu }=\Omega^{\alpha}_{\mu \sigma \nu}-\delta^{\alpha}_{\nu}[\partial^{2}_{\mu \sigma}\psi-\Omega_{\mu \sigma}^{\eta}\partial_{\eta}\psi-{\partial_{\mu}\psi}{\partial_{\sigma}\psi}]\\
  \nonumber +\delta^{\alpha}_{\sigma}[\partial^{2}_{\mu \nu}\psi-\Omega_{\mu \nu}^{\eta}\partial_{\eta}\psi-{\partial_{\mu}\psi}{\partial_{\nu}\psi}].\\
\end{eqnarray}
For a non-light-like interval in Weyl's geometry, we get
\begin{eqnarray}
  \nonumber \widehat{B}^{\alpha}_{\mu \sigma \nu }=B^{\alpha}_{\mu \sigma \nu}+\frac{1}{2}\delta^{\alpha}_{\nu}[\partial^{2}_{\mu \sigma}\omega^{(1)}-\Gamma_{\mu \sigma}^{\eta}\partial_{\eta}\omega^{(1)}-\frac{1}{2}{\partial_{\mu}\omega^{(1)}}{\partial_{\sigma}\omega^{(1)}}]\\
  \nonumber -\frac{1}{2}\delta^{\alpha}_{\sigma}[\partial^{2}_{\mu \nu}\omega^{(1)}-\Gamma_{\mu \nu}^{\eta}\partial_{\eta}\omega^{(1)}-\frac{1}{2}{\partial_{\mu}\omega^{(1)}}{\partial_{\nu}\omega^{(1)}}],\\
\end{eqnarray}
for the symmetric part of the asymmetric geometry and we get
\begin{eqnarray}
  \nonumber \widehat{\Omega}^{\alpha}_{\mu \sigma \nu }=\Omega^{\alpha}_{\mu \sigma \nu}-\frac{1}{2}\delta^{\alpha}_{\nu}[\partial^{2}_{\mu \sigma}\omega^{(1)}-\Omega_{\mu \sigma}^{\eta}\partial_{\eta}\omega^{(1)}-\frac{1}{2}{\partial_{\mu}\omega^{(1)}}{\partial_{\sigma}\omega^{(1)}}]\\
  \nonumber +\frac{1}{2}\delta^{\alpha}_{\sigma}[\partial^{2}_{\mu \nu}\omega^{(1)}-\Omega_{\mu \nu}^{\eta}\partial_{\eta}\omega^{(1)}-\frac{1}{2}{\partial_{\mu}\omega^{(1)}}{\partial_{\nu}\omega^{(1)}}].\\
\end{eqnarray}
for the anti-symmetric part of the asymmetric geometry. 
For a light-like interval in Weyl's geometry, we have
\begin{eqnarray}
  \nonumber \widehat{B}^{\alpha}_{\mu \sigma \nu }=B^{\alpha}_{\mu \sigma \nu}+\delta^{\alpha}_{\nu}[\partial^{2}_{\mu \sigma}\omega^{(2)}-\Gamma_{\mu \sigma}^{\eta}\partial\eta\omega^{(2)}-{\partial\omega^{(2)}}{\partial\sigma \omega^{(2)}}]\\
  \nonumber  -\delta^{\alpha}_{\sigma}[\partial^{2}_{\mu \nu}\omega^{(2)}-\Gamma_{\mu \nu}^{\eta}\partial\eta\omega^{(2)}-{\partial\mu\omega^{(2)}}{\partial\nu \omega^{(2)}}],\\
\end{eqnarray}
for the symmetric part of the asymmetric geometry and we have
\begin{eqnarray}
  \nonumber \widehat{\Omega}^{\alpha}_{\mu \sigma \nu }=\Omega^{\alpha}_{\mu \sigma \nu}-\delta^{\alpha}_{\nu}[\partial^{2}_{\mu \sigma}\omega^{(2)}-\Omega_{\mu \sigma}^{\eta}\partial\eta\omega^{(2)}-{\partial\omega^{(2)}}{\partial\sigma \omega^{(2)}}]\\
  \nonumber +\delta^{\alpha}_{\sigma}[\partial^{2}_{\mu \nu}\omega^{(2)}-\Omega_{\mu \nu}^{\eta}\partial\eta\omega^{(2)}-{\partial\mu\omega^{(2)}}{\partial\nu \omega^{(2)}}].\\
\end{eqnarray}
for the anti-symmetric part of the asymmetric geometry.
We can construct several geometries. For example, if, as in Einstein-Cartan geometry \cite{5}, we choose the symmetric part of (2.1) as the Christoffel symbol plus a symmetric term $ D^{\eta}_{\mu \nu} $, we will have a link between asymmetric geometry and Einstein's geometry.
\newline
There is a possibility of assuming $ \Gamma_{\mu \sigma}^{\eta}={W}_{\mu \nu}^{\eta} $ in (2.2), in which $ {W}_{\mu \nu}^{\eta} $ is given by (2.15) and, for a non-light-like interval in which (2.23) is obeyed, we have
\begin{equation}
\Gamma_{\mu \sigma}^{\eta}={W}_{\mu \nu}^{\eta}=C_{\mu \nu}^{\eta}+{\delta}_{\mu}^{\eta}\partial_{\nu}\omega^{(1)}+{\delta}_{\nu}^{\eta}\partial_{\mu}\omega^{(1)}-g_{\mu \nu}g^{\eta\sigma}\partial_{\sigma}\omega^{(1)},
\end{equation}
and, for a light-like interval, with the condition (2.28),
\begin{equation}
\Gamma_{\mu \sigma}^{\eta}={W}_{\mu \nu}^{\eta}=C_{\mu \nu}^{\eta}+{\delta}_{\mu}^{\eta}\partial_{\nu}\omega^{(2)}+{\delta}_{\nu}^{\eta}\partial_{\mu}\omega^{(2)}-g_{\mu \nu}g^{\eta\sigma}\partial_{\sigma}\omega^{(2)}.
\end{equation}
Then, the symmetric part of the asymmetric connection goes from (2.59) to (2.53) and from (2.61) to (2.54). This asymmetric geometry is composed of two parts, one is two Weyl symmetric geometries and  another is two anti-symmetric geometries. 
\newpage
\renewcommand{\theequation}{\thesection.\arabic{equation}}
\section{First-Order Linear Partial Differential Equation}
\setcounter{equation}{0}
 $         $

Equations (2.22) and (2.25) are two first-order linear partial differential equations and will be solved in this section. Although they represent different intervals of spacetime with different solutions for different values of k, we will consider them in a compact form given below.
\newline
There is more than one method for solving first-order linear partial differential equations. However, for equation (2.22) or (2.25), the most appropriate is the method of change of variables, \cite{3}.
\newline
Let us rewrite (2.22) and (2.25) as follows 
\begin{equation}
h+\frac{\partial{h}}{\partial{z^{\sigma}}}z^{\sigma}=e^{k\Omega},
\end{equation}
with $k=1,2$.
\newline
From this, we get the homogeneous equation
\begin{equation}
\frac{\partial{h}}{\partial{z^{\sigma}}}z^{\sigma}=0,
\end{equation}
which can be put in the characteristic form
\begin{equation}
\frac{dz^{1}}{{z^{1}}}=\frac{dz^{2}}{{z^{2}}}=.....=\frac{dz^{n}}{{z^{n}}},
\end{equation}
with solution given by
\begin{equation}
\Pi(\frac{z^{1}}{{z^{n}}}, \frac{z^{2}}{{z^{n}}}, ....., \frac{z^{n-1}}{{z^{n}}}),
\end{equation}
in which $\Pi$ is an arbitrary function and   
\begin{equation}
 u^{i}=\frac{z^{i}}{{z^{n}}}, 
\end{equation}
$(i=1,2,3,...,n-1)$,
\newline
is a basis of solutions of (3.2).
\newline
In general, $\Pi$ is calculated by setting a boundary value problem, known as the Cauchy problem.
\newline
We change the variables $(z^{1},z^{2},...,z^{n})$ to $(z^{1},u^{2}, ....u^{n-1})$ and (3.1) goes to
\begin{equation}
h(z^{1},u^{1},u^{2}, ....u^{n-1})+\frac{\partial{h}}{\partial{z^{1}}}z^{1}=e^{k\Omega}.
\end{equation}
The solution of (3.6) is given by
\begin{equation}
h(z^{1},u^{1},u^{2}, ....u^{n-1})=\frac{1}{z^{1}}[\Pi(z^{1},u^{1},u^{2}, ....u^{n-1})+\int{e^{k\Omega}}dz^{1}].
\end{equation}
After the integration, we recover the initial variables $(z^{1},z^{2},...,z^{n})$ and the solution assumes the following form 
\begin{equation}
h(z^{1},z^{2},....z^{n})=\frac{1}{z^{1}}[\Pi(\frac{z^{1}}{{z^{n}}}, \frac{z^{2}}{{z^{n}}}, ....., \frac{z^{n-1}}{{z^{n}}})+\int{e^{k\Omega}}dz^{1}].
\end{equation}
Then, in Weyl's geometry, for a non-light-like interval, we get
\begin{equation}
h_{(1)}(z^{1},z^{2},....z^{n})=\frac{1}{z^{1}}[\Pi(\frac{z^{1}}{{z^{n}}}, \frac{z^{2}}{{z^{n}}}, ....., \frac{z^{n-1}}{{z^{n}}})+\int{e^{\omega^{(1)}}}dz^{1}].
\end{equation}
and, for a light-like interval,
\begin{equation}
h_{(2)}(z^{1},z^{2},....z^{n})=\frac{1}{z^{1}}[\Pi(\frac{z^{1}}{{z^{n}}}, \frac{z^{2}}{{z^{n}}}, ....., \frac{z^{n-1}}{{z^{n}}})+\int{e^{2\omega^{(2)}}}dz^{1}].
\end{equation}
For (3.9) and (3.10), we have, for each Cauchy problem of h and its first partial derivatives, a possible set of $\Pi $, $\omega^{(1)}$ and $\omega^{(2)}$, compatible with each Cauchy problem.
\newpage
\renewcommand{\theequation}{\thesection.\arabic{equation}}
\section{Integrability Conditions}
\setcounter{equation}{0}
    $           $

An infinitesimal variation of a vector $ \xi^{\eta} $, is given by
\begin{equation}
d\xi^{\eta}=-{W}_{\mu \nu}^{\eta}dx^{\mu}\xi^{\nu}.
\end{equation}
In Weyl's geometry, from an intuitive viewpoint, (2.13) is equivalent to \cite{6}
\begin{equation}
dl=\omega_{\sigma}dx^{\sigma}l,
\end{equation}
in which $ dl $ is an infinitesimal displacement on a path.
\newline
The integrability condition is given by 
\begin{equation}
\omega_{\mu,\nu}=\omega_{\nu,\mu},
\end{equation}
which is equivalent to
\begin{equation}
\oint{\frac{dl}{l}}=\oint{\omega_{\sigma}dx^{\sigma}}=0.
\end{equation}
If (4.4) is obeyed and $\omega_{\sigma}$ is the gradient of one arbitrary scalar field, then Weyl's geometry is usually presented as reducible to Riemannian geometry. It was developed considering $ \omega_{\sigma} $ as an arbitrary vector field. However, in \cite{1}, it was shown that $ \omega_{\sigma} $ is the gradient of a non-arbitrary scalar field $\omega$ that obeys (2.19) and (2.22).
\newline
We use the following integrability condition
\begin{equation}
\partial^{2}_{\mu\nu}\omega=\partial^{2}_{\nu\mu}\omega,
\end{equation}
which is equivalent to
\begin{equation}
\oint{\frac{dl}{l}}=\oint{{\partial_{\sigma}\omega}dx^{\sigma}}=0.
\end{equation}
The condition (4.6) is correct, but we need to consider two scalar fields, 
$ \omega^{(1)} $ and $ \omega^{(2)} $ respectively obeying (2.19) and (2.22).
\newline
From conformal transformation given by (2.39) and (2.41), Weyl's geometries are always reducible to Riemannian geometry. We call this the equivalence between Einstein and Weyl geometries.
\newline 
If a scalar function is analytic at a point and in its neighbourhood, then (4.5) is an integrability condition.
\newline
Null and massive particles interact locally in an Einstein spacetime. To preserve this and ensure the equivalence between Weyl's and Einstein's geometries, we need (2.53) to be a mapping that sends a massive particle from Einstein's geometry to a Weyl's geometry of non-light-like intervals and (2.54) to be a mapping  sending a null particle from Einstein's geometry to another Weyl's geometry of light-like intervals. If (4.6) is obeyed, then (2.53) and (2.54) can be put as nearly the same Weyl spacetime (see definition below). However, condition (4.6) is not enough. For both particles of Einstein's spacetime to be related to a nearly the same Weyl spacetime, we need two conditions besides (4.6). 
\newline
As an example, consider $ \omega^{(1)} $ and $ \omega^{(2)} $ given by
\begin{equation}
\omega^{(1)}=f_{(1)}+A_{1}sin\alpha_{1}+B_{1}cos\alpha_{1},
\end{equation}
\begin{equation}
2\omega^{(2)}=f_{(2)}+A_{2}sin\alpha_{2}+B_{2}cos\alpha_{2},
\end{equation}
and
\begin{eqnarray}
\nonumber \omega^{(1)}-2\omega^{(2)}=f_{(1)}-f_{(2)}+A_{1}sin\alpha_{1}-\\
\nonumber +B_{1}cos\alpha_{1}-A_{2}sin\alpha_{2}-B_{2}cos\alpha_{2} ,\\
\end{eqnarray}
in which $A_{1}, A_{2}, B_{1}, B_{2}$ are constants,  $\alpha_{i}=K_{(i)\mu}x^{\mu}$, and $f_{(i)}$  are non-periodic functions. 
\newline
We replace (4.7) in (2.39) and (4.8) in (2.41). Then, part of the metrics  $ G^{(1)}_{\mu\nu} $ and  $ G^{(2)}_{\mu\nu} $  oscillate.
\newline
Consider now $ f_{(1)}=f_{(2)} $ and substitute (4.9) in (4.6). If we choose periodic functions that produce stationary waves, (4.6) will also be obeyed.
\newline
Then we can extend (2.56) to 
\begin{equation}
\omega^{(1)}-2\omega^{(2)}=stationary waves.
\end{equation}
In general, stationary waves and resonances are linear phenomena.
\newline
Condition (4.10) implies that null and massive particles in one Einstein's spacetime can be related to nearly the same Weyl's spacetime for some points of spacetime.
\newline
If $A_{1}, A_{2}, B_{1}, B_{2}$ are non-infinitesimal constants, the condition (4.10) is not enough, since geodesics and curvatures may be very different for $ \omega^{(1)}$ and $\omega^{(2)}$. Then (4.10) should be replaced by
\begin{equation}
\omega^{(1)}-2\omega^{(2)}=(infinitesimal constant).(linear). (stationary waves).
\end{equation}
\newline
Then, massive particles of one Einstein's spacetime will be related to (2.53) (non-light-like interval) and null particles to (2.54) (light-like interval) in an infinity of points of spacetime. In other words, in the interval $ x^{\mu} $ and $ x^{\mu}+dx^{\mu} $, massive particles will be in (2.53) a number infinite of times and null particles in (2.54) a number infinite of times.
\newline
We now define "nearly the same Weyl spacetime". It is two Weyl's geometries given by (2.53) (non-light-like interval) and (2.54) (light-like interval)and obeying (4.11) and (4.6).
\newline
For non-light-like and light-like intervals, condition (4.11) is enough so that geodesics and curvatures of the two spaces, (2.53) and (2.54), differ by small fluctuations. Moreover, null and massive particles coexist in nearly the same Weyl's spacetime and are also particles in Einstein's spacetime.  
\newline
For a set of different interacting particles in one Einstein's spacetime, we can associate a nearly the same Weyl's geometry for each particle.
\newline
It is clear that (4.10) and (4.11) obey a classical wave equation.
\renewcommand{\theequation}{\thesection.\arabic{equation}}
\section{Transverse Waves in a Flat Oscillating Spacetime}
\setcounter{equation}{0}
    $           $
Longitudinal waves can also be stationary and, together with stationary transverse waves, can be present in (4.11). Then, we can have longitudinal and transverse stationary waves as spacetime oscillations. From the classical viewpoint, some phenomena are governed by transverse waves, such as electromagnetic fields in a vacuum.
\newline
In this paper, we will use only stationary linear transverse waves, excluding longitudinal waves in (4.11).
\newline
Let us consider Hamilton's equation for a classical particle in a flat oscillating spacetime \cite{7}
\begin{equation}
H=\frac{1}{2m}[(P_{x})^{2}+(P_{y})^{2}+(P_{z})^{2}+(P_{x})^{2}]+V(x,y,z),
\end{equation}
in which $V(x, y, z)$ is a conservative potential.
\newline
Hamilton's phase function is defined by
\begin{equation}
S=W(x,y,z)-Et,
\end{equation}
in which E is energy.
\newline
The relationship between momenta and the phase function S is given by the gradient of
S,
\begin{equation}
 P_{x}=\frac{\partial{S}}{\partial{x}}, P_{y}=\frac{\partial{S}}{\partial{y}}, P_{z}=\frac{\partial{S}}{\partial{z}},
\end{equation}
with the associated Hamilton-Jacobi equation,
\begin{equation}
\frac{1}{2m}[(\frac{\partial{S}}{\partial{x}})^{2}+(\frac{\partial{S}}{\partial{y}})^{2}+(\frac{\partial{S}}{\partial{z}})^{2}]+V(x,y,z)+\frac{\partial{S}}{\partial{t}}=0.
\end{equation}
The function S defines phase surfaces from the condition
\begin{equation}
S=W(x,y,z)-Et=constant,
\end{equation}
from which the Hamilton-Jacobi equation will be
\begin{equation}
\frac{1}{2m}[(\frac{\partial{W}}{\partial{x}})^{2}+(\frac{\partial{W}}{\partial{y}})^{2}+(\frac{\partial{W}}{\partial{z}})^{2}]+V(x,y,z)=E.
\end{equation}
Let us consider two phase surfaces separated by an infinitesimal time interval. The
particle momentum is given by the gradient of S and the phase velocity will be
\begin{equation}
V_{ph}=\frac{E}{|grad W|}
\end{equation}
From (5.6), we have the gradient of W
\begin{equation}
(grad W)^{2}=(\frac{\partial{W}}{\partial{x}})^{2}+(\frac{\partial{W}}{\partial{y}})^{2}+(\frac{\partial{W}}{\partial{z}})^{2}= 2m(E-V(x,y,z)).
\end{equation}
Using (5.8) in (5.7), we obtain
\begin{equation}
(V_{ph})^{2}=\frac{{E}^{2}}{2m(E-V)}.
\end{equation}
A totally transverse wave in a material medium has very strange properties, such as infinite rigidity in the direction of propagation. However, for a flat oscillating spacetime, this difficulty does not exist.
\newline
We now present a transverse wave equation describing the motion of a particle in a flat oscillating spacetime. 
\newline
Consider that the particle is not dragged by waves of spacetime and also does not drag waves. Thus, the particle velocity, phase velocity and wave group velocity will be equal to each other and therefore a wave equation for a transverse wave that describes the motion of the particle will be given by
\begin{equation}
\triangledown^{2}\Theta-\frac{2m(E-V)}{{E}^{2}}\partial^{2}_{t}\Theta=0.
\end{equation}
\newline
If we put $ \Theta $ with an infinitesimal amplitude, we have an equivalence with $A_{1}, A_{2}, B_{1}, B_{2}$,  which are infinitesimal constants in (4.6), and the condition (4.11) will be obeyed. This allows the transverse wave to propagate in Einstein's geometry and only in Weyl's geometry of (2.53)(non-light-like interval), because we have a wave equation describing the motion of a massive particle.
\newline
If a massive particle moves in a trajectory that obeys the condition (4.11) and (4.6), then we have a superposition of stationary waves. If an external field interacts with the particle and modifies its trajectory, in Einstein's geometry, it is possible for the spacetime oscillations to be perturbed and lost their stationary nature, violating (4.11) and (4.6). However, it is also possible for the new trajectory to obeys conditions (4.11) and (4.6) with different sets of stationary waves on old and new trajectories.
\newline
Energy conservation occurs in Bohr's hydrogen model and Schrodinger's quantum mechanics. From Bohr's hydrogen atom and Schrodinger's quantum mechanics, we have a conservative system, where the transition between allowed orbits involves discrete energy variation proportional to the difference in frequencies between the stationary waves of the two orbits. This difference is emitted as a quantized photon. The photon emission or quantum leap is seen as a null particle in the geometry (2.54). From the mathematical viewpoint, a difference in frequencies between the stationary waves of the two orbits is an integer number, because conditions (4.11) and (4.6) need to be obeyed. This suggests that a photon emission from a higher to lower energy orbit obeys this conservation if the sum of the energy of the system is divided as follows, atomic energy living in (2.53), photon energy in (2.54) and atom energy plus photon energy living in Einstein's spacetime. We conclude that a photon in (2.54) is quantized if energy is conserved.
\newline
In (2.8), $(y^{\eta}, z^{\eta})$ are real normal coordinates. Then, $\Omega$ is real in (3.1), but in (5.10) $\Theta$ can be real or complex. This suggests writing a relationship between $\Omega$ and $\Theta$ that includes complex solutions. For this, let us rewrite (3.1)
\begin{equation}
h+\frac{\partial{h}}{\partial{z^{\sigma}}}z^{\sigma}=e^{k\Omega},
\end{equation}
and define 
\begin{equation}
log(h)+\frac{\partial{h}}{\partial{z^{\sigma}}}z^{\sigma})=k\Omega=F(|\Theta|^{2}),
\end{equation}
in which $F(|\Theta|^{2})$ is a function to be determined by an experiment, if possible.
\newline
For a non-light-like interval, we get
\begin{equation}
log(h_{(1)})+\frac{\partial{h_{(1)}}}{\partial{z^{\sigma}}}z^{\sigma}=\omega^{(1)}=F_{(1)}(|\Theta|^{2}),
\end{equation}
and, for a light-like interval,
\begin{equation}
log(h_{(2)}+\frac{\partial{h_{(2)}}}{\partial{z^{\sigma}}}z^{\sigma})=2\omega^{(2)}=F_{(2)}(|\Theta|^{2}).
\end{equation}
Rewriting (5.10) and (3.8), we get, respectively 
\begin{equation}
\triangledown^{2}\Theta-\frac{2m(E-V)}{{E}^{2}}\partial^{2}_{t}\Theta=0,
\end{equation}
\begin{equation}
h(z^{1},z^{2},....z^{n})=\frac{1}{z^{1}}[\Pi(\frac{z^{1}}{{z^{n}}}, \frac{z^{2}}{{z^{n}}}, ....., \frac{z^{n-1}}{{z^{n}}})+\int{e^{k\Omega}}dz^{1}].
\end{equation}
Using (5.12) in (5.16), we have
\begin{equation}
h(z^{1},z^{2},....z^{n})=\frac{1}{z^{1}}[\Pi(\frac{z^{1}}{{z^{n}}}, \frac{z^{2}}{{z^{n}}}, ....., \frac{z^{n-1}}{{z^{n}}})+\int{e^{F(|\Theta|^{2})}}dz^{1}].
\end{equation}
For a non-light-like interval, we get
\begin{equation}
h_{(1)}(z^{1},z^{2},....z^{n})=\frac{1}{z^{1}}[\Pi(\frac{z^{1}}{{z^{n}}}, \frac{z^{2}}{{z^{n}}}, ....., \frac{z^{n-1}}{{z^{n}}})+\int{e^{F_{(1)}(|\Theta|^{2})}}dz^{1}],
\end{equation}
and, for a light-like interval,
\begin{equation}
h_{(2)}(z^{1},z^{2},....z^{n})=\frac{1}{z^{1}}[\Pi(\frac{z^{1}}{{z^{n}}}, \frac{z^{2}}{{z^{n}}}, ....., \frac{z^{n-1}}{{z^{n}}})+\int{e^{F_{(2)}(|\Theta|^{2})}}dz^{1}].
\end{equation}
In Weyl's geometry, for a non-light-like interval, (2.35) will be
\begin{equation}
\widehat{X}^{\eta}=e^{-\omega^{(1)}}X^{\eta}=e^{-F_{(1)}(|\Theta|^{2})}X^{\eta},
\end{equation}
and, for a light-like interval, 
\begin{equation}
\widehat{X}^{\eta}=e^{-2\omega^{(2)}}X^{\eta}=e^{-F_{(2)}(|\Theta|^{2})}X^{\eta}.
\end{equation}
\newline
Although $\Pi(\frac{z^{1}}{{z^{n}}}, \frac{z^{2}}{{z^{n}}},..., \frac{z^{n-1}}{{z^{n}}})$ is arbitrary, it can be known by considering the Cauchy problem. The Cauchy problem needs to obey the condition in which h, given in (3.1), is finite in the origin of normal coordinates. 
\newline
As seen in \cite{1}, equations of type (3.1) are invariant under a transformation from local coordinates        $ x^{\eta}$ to normal coordinates $z^{\eta}$. Then, (5.20) and (5.21) are sufficiently general solutions.
\section{Concluding Remarks}
\setcounter{equation}{0}
In this paper, we presented and analysed important properties of non-light-like and light-like intervals. We showed that there are infinite solutions for scalar fields. We analysed the integrability condition in Weyl's geometry and concluded that the equivalence between Einstein's and Weyl's geometries is compatible with the hypothesis of spacetime oscillations, forming linear and stationary waves at each point of spacetime. We see that (4.6) is also obeyed if the periodic functions produce stationary waves. We hypothesized that these waves do not drag particles and are not dragged by them, so that the motion of particles is governed by these waves and external fields. We constructed an equation for these transverse waves and performed a qualitative application in a hydrogen atom, which, from the qualitative viewpoint, agrees with Bohr's hydrogen model and Schrodinger's quantum mechanics. 
 
\end{document}